\documentclass[aps,twocolumn,amsfonts,amssymb,amsmath]{revtex4}
\usepackage{epsfig, graphicx, color, amsmath, amsthm, amssymb}
\def\ket #1{\vert #1\rangle}
\def\bra #1{\langle #1\vert}
\def\ketbra #1#2{\ket{#1}\!\bra{#2}}

\def\abs #1{\lvert #1\rvert}
\DeclareMathOperator{\tr}{Tr}
\newcommand{\beq}{\begin{equation}}
\newcommand{\eeq}{\end{equation}}

\newlength{\commentslength}

\newtheorem{theorem}{Theorem}  

\newtheorem{corollary}{Corollary}

\begin{document}

\author{Ben W. Reichardt}
\email{breic@cs.berkeley.edu}
\thanks{Research supported in part by NSF ITR Grant CCR-0121555, and ARO Grant DAAD 19-03-1-0082.}
\affiliation{UC Berkeley}

\title{Improved magic states distillation for quantum universality}

\begin{abstract}
Given stabilizer operations and the ability to repeatedly prepare a single-qubit mixed state $\rho$, can we do universal quantum computation? 
As motivation for this question, ``magic state" distillation procedures can reduce the general fault-tolerance problem to that of performing fault-tolerant stabilizer circuits.

We improve the procedures of Bravyi and Kitaev in the Hadamard ``magic" direction of the Bloch sphere to achieve a sharp threshold between those $\rho$ allowing universal quantum computation, and those for which any calculation can be efficiently classically simulated. 
As a corollary, the ability to repeatedly prepare any pure state which is not a stabilizer state (e.g., any single-qubit pure state which is not a Pauli eigenstate), together with stabilizer operations, gives quantum universality. 
It remains open whether there is also a tight separation in the so-called T direction.
\end{abstract}

\maketitle

\section{Introduction}

In ``magic states distillation," introduced by Bravyi and Kitaev in \cite{BravyiKitaev04}, we try to achieve universal quantum computation using only Clifford group unitaries, preparation and measurement in the computational basis $\ket{0}$, $\ket{1}$, and the ability to prepare a given single-qubit mixed state $\rho$.

\begin{figure}
\includegraphics*[bb=21 652 156 786,scale=1]{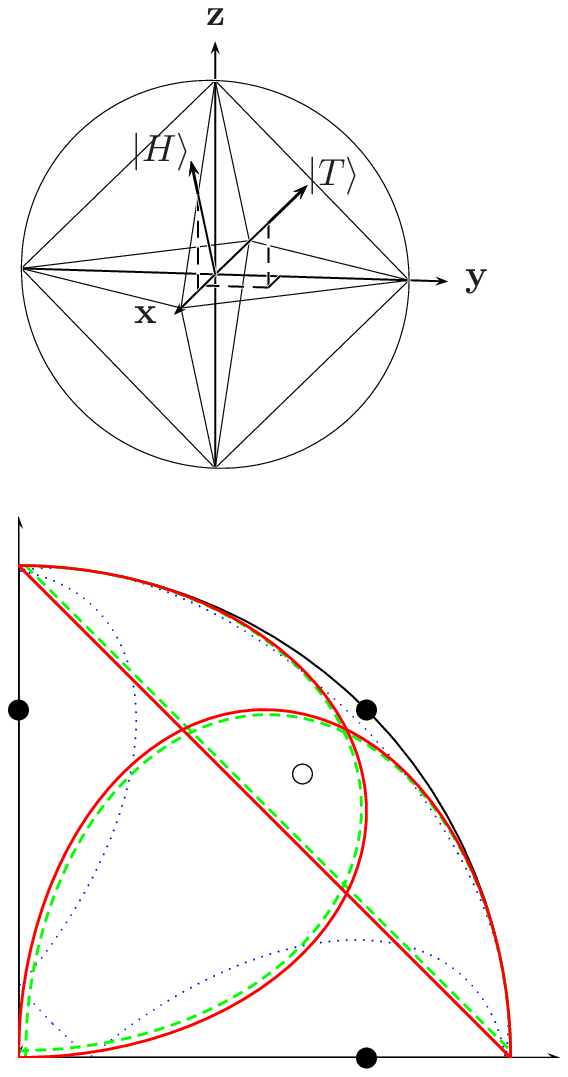}
\caption{The Bloch sphere, the octahedron $O$, and the magic states $\ket{H}$ and $\ket{T}$.} \label{f:blochsphere}
\end{figure}

If $\rho$, considered as a point in the Bloch sphere of Fig.~\ref{f:blochsphere}, lies within $O$ the octahedral closed convex hull of the six eigenvectors of the Pauli operators $X$, $Y$ and $Z$, then the calculation is classically simulable by the Gottesman-Knill theorem.
Bravyi and Kitaev show universality if $\rho$ is one of certain pure states: either Hadamard eigenstates (midway between $X$ and $Z$ eigenstates on the Bloch sphere) and symmetrical states under the symmetries of the octahedron $O$ (i.e. radially out from the midpoints of the edges of $O$), or states symmetrical to $T$ eigenstates (radially out from the centers of the faces of $O$).
Here the operator $T$ conjugates the Pauli operators as $X \rightarrow Y \rightarrow Z \rightarrow X$, a $2\pi/3$ rotation (order $3$) on the Bloch sphere.  (Prior to \cite{BravyiKitaev04},  \cite{KnillLaflammeZurek98} had shown universality with the Hadamard eigenstate, and see \cite{BoykinMorPulverRoychowdhuryVatan00}.)

Bravyi and Kitaev show that the restricted set of operations can be used to purify $\rho$ in the $H$ and $T$ ``magic" directions.  Here is a summary of their results in the $T$ direction.  By randomly applying the $T$ gate, we may assume that $\rho$ is a mixture of the $e^{2\pi i/3}$ and $e^{-2\pi i/3}$ eigenstates with probabilities $1-p$ and $p$, respectively.  Bravyi and Kitaev give a method, based on decoding a well-known 5-qubit code, for taking five copies of $\rho$ with error $p < \tfrac{1}{2}(1-\sqrt{3/7})$, and (with some probability) producing a mixture in the $T$ direction with smaller error $\tfrac{t^5+5t^2}{1+5t^2+5t^3+t^5}$, where $t \equiv \tfrac{p}{1-p}$.  Recursively applying the method allows the reduction of $p$ exponentially fast in $n$ the number of prepared copies of $\rho$ ($p_{\text{out}}(n,p) \sim (5p)^{n^{1/\log_2 30}}$).  Geometrically, in the Bloch sphere, the threshold condition is that $\rho$ lies beyond a plane parallel to a face of $O$ at distance $\sqrt{3/7} \approx 0.655$ from the origin.  It is an interesting open question whether universal quantum computation is possible for states $\rho$ between this plane and the face of $O$ (which is distance $1/\sqrt{3} \approx 0.577$ from the origin, or $p = \tfrac{1}{2}(1-1/\sqrt{3})$).  In terms of the parameter $\rho$, is there a tight separation between those mixed states allowing only classically-simulable computation and those allowing universal quantum computation?

The $H$ direction bisects an edge of the octahedron $O$.  Again, we may assume that $\rho$ is a mixture of the $+1$ and $-1$ eigenstates of the Hadamard gate $H$, with probabilities $1-p$ and $p$.  Bravyi and Kitaev give a particular 15-qubit code.  They take 15 copies of $\rho$, and apply the decoding circuit for their code.  They reject if any errors are detected.  After one successful iteration, the output error is
\beq
p_{\text{out}} = \frac{1-15(1-2p)^7+15(1-2p)^8-(1-2p)^{15}}{2(1+15(1-2p)^8)} \enspace , \nonumber
\eeq
giving an error threshold of about $14.148\%$ below which $H$ eigenstates can be distilled.
Interestingly, Knill \cite{Knill04schemes} had earlier independently shown an apparently quite different method which however achieves the exact same $p_{\text{out}}$.  Knill's method is based on using fourteen copies of $\rho$ to apply a faulty logical controlled-Hadamard to the 7-bit Steane/Hamming code.
We analyzed Knill's method and found that starting with the 7 bits in an encoding of a fifteenth copy of $\rho$ gave the best results --
surprisingly, in fact this distillation method is exactly equivalent to Bravyi and Kitaev's method (in terms of output error in terms of $p$; the acceptance probability is actually $2^{14}$ times smaller).

In this paper, we will show that in fact the separation between classical simulability and universal quantum computation is tight in the $H$ direction.
\begin{theorem}\label{t:hdistill}
Let $F_T(\rho)$ be the maximum fidelity between $\rho$ and an $H$-type magic state, i.e.
$$
F_T(\rho)=\max_U \sqrt{\bra{H}U^\dagger \rho U \ket{H}} \enspace ,
$$
where $U$ ranges over the symmetries of the octahedron $O$.
Preparation of $\rho$ together with Clifford group operations and Pauli eigenstate preparation and measurement allows universal quantum computation whenever
$$
F_T(\rho) > F_H^* \overset{\mathrm{def}}{=} \left[\frac{1}{2}\left(1+\sqrt{\frac{1}{2}}\right)\right]^{\frac{1}{2}} \approx 0.924 \enspace .
$$
\end{theorem}
We show that decoding either the Steane 7-qubit code or the Golay 23-qubit code gives the result.  Other CSS codes might also work, as we'll discuss in Appendix~\ref{s:moregeneralCSS}.  Fig.~\ref{f:blochquadrant} shows the improvement in the threshold on $\rho$ for $H$-type distillation using our new procedure.
As an immediate consequence of this theorem, we obtain:

\begin{corollary} \label{t:purestatecor}
Preparation of any single-qubit pure state excepting the six Pauli eigenstates, together with Clifford group operations and Pauli eigenstate preparation and measurement, allows universal quantum computation.
\end{corollary}

(Compare to the universality results of \cite{Shi02}.)  

\begin{figure}[!t] 
\includegraphics*[bb=20 482 177 640,scale=1]{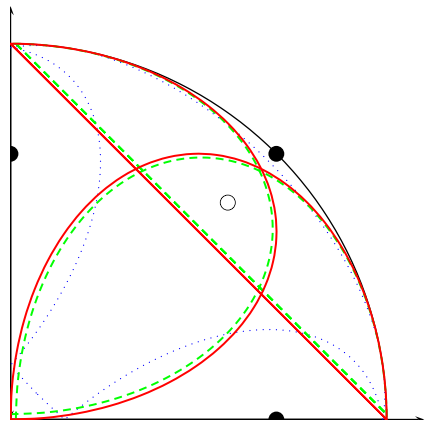}
\caption{Looking down on one quadrant of the Bloch sphere.  The small empty and filled circles respectively indicate $T$-type and $H$-type magic states.  The dotted blue projected circles give the intersections with the unit sphere of the planes beyond which $T$ distillation is possible.  The dashed green circles similarly show the limits of Bravyi and Kitaev's $H$ distillation procedure, and the red circles the limits of our new $H$ distillation procedure.  Examining the origin at lower left, one sees that all pure states excepting the Pauli eigenstates can be distilled to $\ket{H}$.}\label{f:blochquadrant}
\end{figure}

In Appendix~\ref{s:multiqubitpurestatecor}, we show that any multi-qubit pure state which is not a stabilizer state can be reduced to a single-qubit pure state which is not a Pauli eigenstate using Clifford group operations and postselected Pauli measurements, proving:

\begin{corollary} \label{t:multiqubitpurestatecor}
Preparation of any pure state which is not a stabilizer state, together with Clifford group operations and Pauli eigenstate preparation and measurement, allows universal quantum computation.
\end{corollary}



Besides the theoretical interest of the magic states distillation problem, there is a more practical interest.  This is how Knill came upon the same problem.  In error-correction threshold schemes, Clifford group operations are typically easiest to analyze and simulate because they take Pauli product errors to other Pauli product errors.  However, analyzing only Clifford group operations will not give a threshold for universal quantum computation.  Knill's solution is to prepare an encoded Bell pair, decode one half of it (throwing away the entire pair if any errors are detected), then teleport a single bit state $\rho$ into the encoding.  If the errors in $\rho$, the Bell pair, and the Bell measurement used for teleportation are small enough, then the encoded $\rho$ can be distilled to a magic state.  Our improved distillation procedure does not increase Knill's threshold in \cite{Knill04analysis}, however, because this step is not the bottleneck.
A related problem is determining the power of error-prone gate sets \cite{HarrowNielsen,VirmaniHuelgaPlenio}.

We give two proofs of Theorem~\ref{t:hdistill}.
In Sections~\ref{s:Steane} and~\ref{s:Golay}, we show the thresholds for the $[[7,1,3]]$ Steane code and the $[[23,1,7]]$ Golay code.



\section{Steane $[[7,1,3]]$ code} \label{s:Steane}

In this distillation procedure, we take seven copies of
$$
\rho = \left(\begin{matrix}
\rho_{00} & \rho_{01} = \bar{\rho}_{10} \\
\rho_{10} & \rho_{11} = 1 - \rho_{00}
\end{matrix}\right)
\enspace ,
$$
and decode the Steane $[[7,1,3]]$ code, rejecting and starting over if any errors are detected.  The 7-bit Steane code has stabilizer generators
\begin{gather*}
IIIXXXX,IXXIIXX,XIXIXIX,\\
IIIZZZZ,IZZIIZZ,ZIZIZIZ \enspace ,
\end{gather*}
and its logical $X$ and logical $Z$ operations are simply transverse $X$ and $Z$, respectively.  Let $S$ be the set of even-weight codewords to the classical 7-bit Hamming code:
$$
S = \left\{ \begin{matrix}0^7, 0001111, 0110011, 0111100, 1010101, \\ 1011010, 1100110, 1101001 \end{matrix} \right\} \enspace .
$$
Then the $\pm1$ logical $Z$ eigenstates $\ket{0_L}$, $\ket{1_L}$ are given by
\begin{gather*}
\sqrt{8} \ket{0_L} = \ket{0_L^{(0)}} + \ket{0_L^{(4)}} \enspace \,\\
\sqrt{8} \ket{1_L} = \ket{1_L^{(7)}} + \ket{1_L^{(3)}} \enspace ,
\end{gather*}
where $\ket{0_L^{(0)}} = \ket{0^7}$, $\ket{0_L^{(4)}} = \sum_{a \in S - 0} \ket{a}$, and $\ket{1_L^{(7)}} = X_L \ket{0_L^{(0)}} = X^{\otimes 7} \ket{0_L^{(0)}}$, $\ket{1_L^{(3)}} = X_L \ket{0_L^{(4)}}$.  (The superscript gives the weight of the classical codewords summed over.)

We compute 
\begin{eqnarray*}
\bra{0_L} \rho^{\otimes 7} \ket{0_L}
&=&
\frac{1}{8} \left(\begin{split}
\bra{0^7} \rho^{\otimes 7} \ket{0^7}
+
2 \Re \bra{0^7} \rho^{\otimes 7} \ket{0_L^{(4)}}\\
+
\bra{0_L^{(4)}} \rho^{\otimes 7} \ket{0_L^{(4)}}
\end{split}\right) \\
&=&
\tfrac{1}{8}(
\rho_{00}^7+2 \cdot 7 \Re \rho_{00}^3 \rho_{01}^4+\bra{0_L^{(4)}} \rho^{\otimes 7} \ket{0_L^{(4)}}
) \enspace .
\end{eqnarray*}
where we have applied that $\bra{0^7} \rho^{\otimes 7} = (\rho_{00} \bra{0} + \rho_{01} \bra{1})^{\otimes 7}$.

Then use
\begin{eqnarray*}
\bra{0_L^{(4)}} \rho^{\otimes 7} \ket{0_L^{(4)}}
&=&
\sum_{a \in S-0} \bra{a}\rho^{\otimes7}\ket{a}
+
\sum_{\substack{a,b\in S-0\\a\neq b}} \bra{a}\rho^{\otimes7}\ket{b} \\
&=&
7 \rho_{00}^3\rho_{11}^4 + 7 \cdot 6 \rho_{00}^1\rho_{01}^2\rho_{10}^2\rho_{11}^2
.
\end{eqnarray*}
The first term is because each element of $S - 0^7$ has Hamming weight 4.  The second term is because any two distinct elements of $S - 0^7$ share a 0 in one position, share 1's in two positions, and differ (one being 0, the other 1) in four positions.
Substitute back in to get
\begin{multline*}
\bra{0_L} \rho^{\otimes 7} \ket{0_L} = \tfrac{1}{8} ( \rho_{00}^7 + 7 \rho_{00}^3(\rho_{01}^4+\rho_{10}^4)\\+7\rho_{00}^3\rho_{11}^4+7 \cdot 6\rho_{00} \abs{\rho_{01}}^4 \rho_{11}^2 ) \enspace .
\end{multline*}

We can then use that the terms of $\ket{1_L}$ are simply the bitwise complements of terms of $\ket{0_L}$ to obtain
\begin{widetext}
\begin{eqnarray*}
\bra{1_L} \rho^{\otimes 7} \ket{1_L} &=&
\tfrac{1}{8} ( \rho_{11}^7 + 7\rho_{11}^3(\rho_{01}^4+\rho_{10}^4) + 7\rho_{00}^4\rho_{11}^3 + 7 \cdot 6 \rho_{00}^2 \abs{\rho_{01}}^4 \rho_{11} ) \\
\bra{0_L} \rho^{\otimes 7} \ket{1_L} &=&
\tfrac{1}{8} ( \rho_{01}^7 + 7\rho_{01}^3(\rho_{00}^4+\rho_{11}^4) + 7\rho_{10}^4\rho_{01}^3 + 7 \cdot 6 \rho_{00}^2 \rho_{01}\rho_{10}^2 \rho_{11}^2 ) \enspace .
\end{eqnarray*}
\end{widetext}

\begin{figure}
\includegraphics*[bb=92 3 426 210,scale=.70]{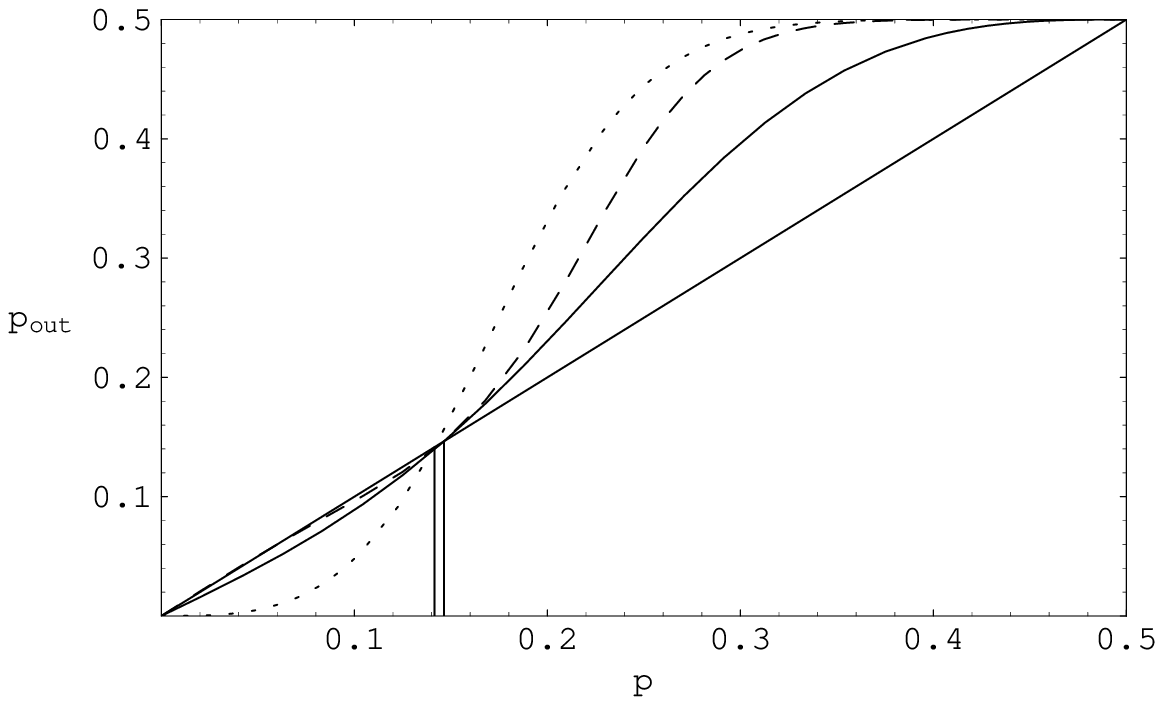}
\includegraphics*[bb=92 3 426 210,scale=.70]{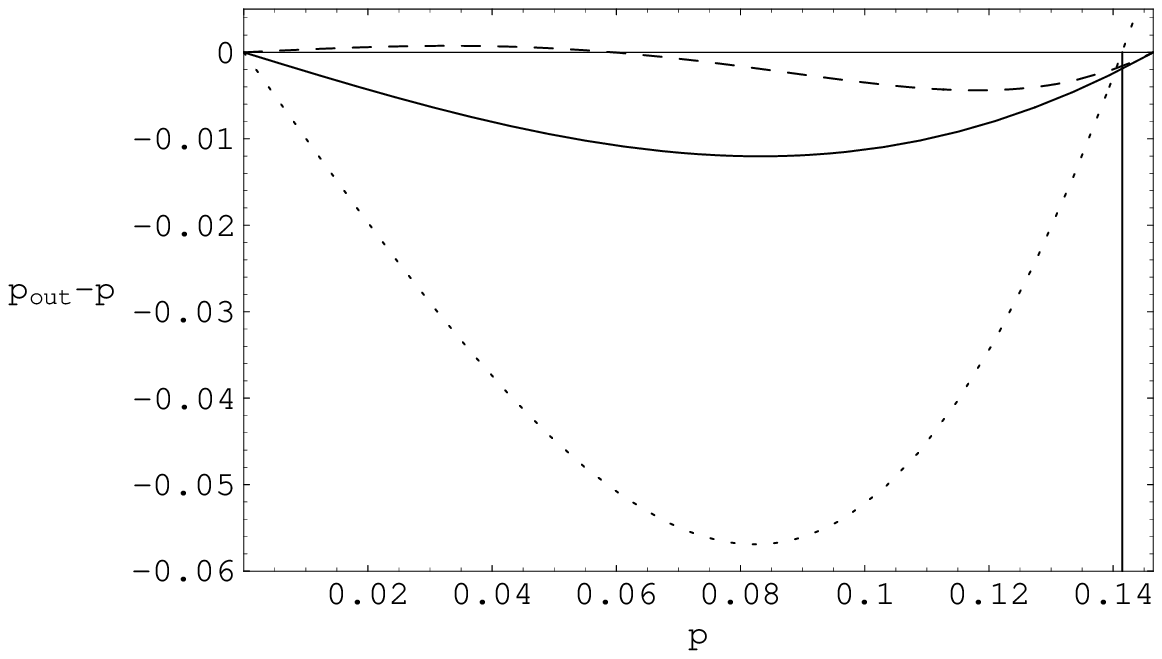}
\caption{The final error probability $p_{out}$ as a function of the initial error probability $p$ for the $H$-type distillation.  The dotted curve shows the effectiveness of the distillation procedures of Knill, and Bravyi and Kitaev.  The dashed curve is for the Golay code, and the solid curve for the Steane code.  Below, we plot $p_{out}-p$ for $p$ beneath the threshold of $\tfrac{1}{2}(1-1/\sqrt{2})$.  Notice that $\ket{H}$ itself is only an unstable fixed point for the Golay code distillation procedure, and there is a stable intermediate fixed point.}\label{f:thresholdplot}
\end{figure}

\begin{figure}
\includegraphics*[bb=92 3 426 210,scale=.70]{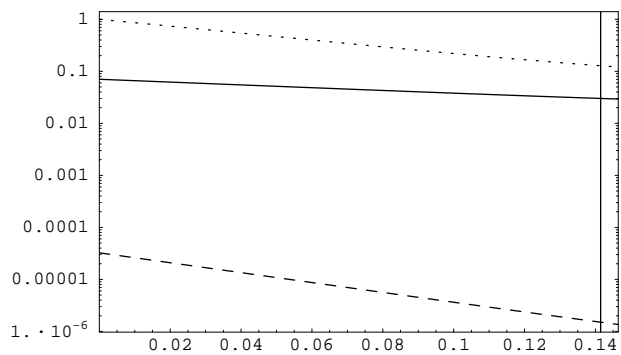}
\caption{The decoding acceptance probability as a function of the initial error probability $p$ for the $H$-type distillation.  The dotted curve is for the distillation procedure of Bravyi and Kitaev; the dashed curve is for the Golay code; and the solid curve for the Steane code.  The new methods do not work as well for low values of $p$.  Using the Golay code is always inferior to using the Steane code.}\label{f:acceptanceprobs}
\end{figure}

By symmetry we may assume $\rho$ lies along a line in the $H$ direction ($x=z$, $y=0$), or $\rho_{10} = \rho_{01} = \tfrac{x}{2}$, $\rho_{00} = \tfrac{1+x}{2}$, $\rho_{11} = \tfrac{1-x}{2}$.  The trace (probability of success) then turns out to be $\tfrac{1}{64}(1+14x^4)$, and the normalized density matrix is
$$
\frac{1}{2+28x^4}
\left(\begin{matrix}
1+7x^3+14x^4+8x^7 & x^3 (7+8x^4) \\
x^3 (7+8x^4)      & 1-7x^3+14x^4-8x^7
\end{matrix}\right)
,
$$
corresponding to $x_{out} = \tfrac{x^3(7+8x^4)}{1+14x^4}$.  Plotting this function shows a threshold at $x = \tfrac{1}{2}$.  Fig.~\ref{f:thresholdplot} shows a plot versus $p = \tfrac{1}{2}(1-\sqrt{2}x)$, and Fig.~\ref{f:acceptanceprobs} plots the efficiency.  This proves Theorem~\ref{t:hdistill}.  Note that in the small $p$ limit, $p_{out} \approx \tfrac{7}{9}p$ (versus $35p^3$ for the procedures of Knill, and Bravyi and Kitaev), so convergence is only polynomially fast in the number of copies of $\rho$ used.  When the error probability is sufficiently small, switching to the 15-qubit code gives an exponential gain in efficiency.

\section{Golay $[[23,1,7]]$ code} \label{s:Golay}

The computations we performed above can be carried out for any CSS code for which the logical $X$ and $Z$ operations are simply transverse $X$ and $Z$, respectively.  The terms that appear are of the form
$$
\sum_{a \in S^{(k)}} \bra{a} \rho^{\otimes n} \ket{a} \qquad \text{or} \qquad \sum_{\substack{a \in S^{(k)}, b \in S^{(l)} \\ a\neq b}} \bra{a} \rho^{\otimes n} \ket{b} \enspace ,
$$
where $S^{(k)}$, $S^{(l)}$ are sets of classical codewords (satisfying the $Z$ parity checks) of weights $k$ and $l$, respectively.  Evaluating the first term requires knowing the weight distribution of the classical code corresponding to the $Z$ stabilizers.  Table~\ref{f:golayweights} gives this distribution for the classical $[23,12,7]$ Golay code.  Evaluating the second term requires knowing the distribution of the Hamming weights of the exor $a\oplus b$ as $a$ and $b$ vary over all codewords of weights $k$ and $l$.  Table~\ref{f:golayinterweights} gives this distribution for the Golay code.

\begin{table}
\caption{The number of codewords of any given weight for the classical $[23,12,7]$ Golay code.  All $2^{12}$ codewords are accounted for.}\label{f:golayweights}
\begin{tabular}{c|c}
Weight & \# codewords \\
\hline
0, 23  & 1    \\
8, 15  & 506  \\
12,11  & 1288 \\
16, 7  & 253
\end{tabular}
\end{table}

\begin{table}
\caption{Varying $a$ over all Golay codewords of weight $k$, and $b$ over all codewords of weight $l$, the $k$,$l$ entry of this table gives the number of times $a \oplus b$ has Hamming weight 0, 8, 12 or 16, in that order.  Only positive, even entries for $k$ and $l$ are shown; zero-weight entries can be determined from Table~\ref{f:golayweights}, and odd-weight entries can be determined since every odd-weight codeword is the bitwise complement of an even-weight codeword.}\label{f:golayinterweights}
\begin{tabular}{c|c|c|c|c}
 & 8 & 12 & 16 \\
\hline
 &506& 0  & 0 \\
\raisebox{-.5ex}{8}&106260&141680& 7590 \\
 &141680&425040& 85008 \\
 &7590&85008& 35420\\
\hline
 &  & 1288  & 0 \\
\raisebox{-.5ex}{12}& & 425040  & 85008 \\
 &  & 1020096 & 212520 \\
 &  & 212520  & 28336\\
\hline
 &  & & 253 \\
\raisebox{-.5ex}{16}& & & 35420 \\
 &  & & 28336 \\
 &  & & 0 \\
\hline
\end{tabular}
\end{table}

To use these tables, note for example that $\bra{0_L} \rho^{\otimes 23} \ket{0_L} = \tfrac{1}{2048} \sum_{k,l \in \{0,8,12,16\}} \bra{k} \rho^{\otimes 23} \ket{l}$, where $\ket{k}$ represents the unnormalized sum over all codewords of weight $k$.  Using the second table we can read off
\begin{multline*}
\bra{8}\rho^{\otimes 23}\ket{12} = 141680 \rho_{00}^9 \rho_{01}^6 \rho_{10}^2 \rho_{11}^6 \\+ 425040 \rho_{00}^7 \rho_{01}^8 \rho_{10}^4 \rho_{11}^4 + 85008 \rho_{00}^5 \rho_{01}^{10} \rho_{10}^6 \rho_{11}^2 \enspace .
\end{multline*}
Here, e.g., the first term comes $a \oplus b$ having weight 8; it is a degree 23 polynomial in entries of $\rho$, for which the exponents of $\rho_{10}$ and $\rho_{11}$ sum to $\abs{a} = 8$, the exponents of $\rho_{01}$ and $\rho_{11}$ sum to $\abs{b} = 12$, and the exponents of $\rho_{01}$ and $\rho_{10}$ sum to $\abs{a \oplus b} = 8$.

Since each odd-weight codeword is the bitwise complement of an even-weight codeword, having computed $\bra{0_L} \rho^{\otimes 23} \ket{0_L}$, $\bra{0_L} \rho^{\otimes 23} \ket{1_L}$ can be obtained by just flipping the second subscripts, taking each occurrence of $\rho_{i,j}$ to $\rho_{i,1-j}$.  To compute $\bra{1_L} \rho^{\otimes 23} \ket{1_L}$, flip both subscripts.

Now substituting for $\rho$ lying on a line in the $H$ direction, we obtain with some algebra that the decoding acceptance probability is
$
(1+1012 x^8 + 2576 x^{12} + 8096 x^{16})/2^{22}
$, and the conditional density matrix has
$$
x_{out} = \frac{x^7(253 + 1288 x^4 + 8096 x^8 + 2048 x^{16})}{1+1012 x^8 + 2576 x^{12} + 8096 x^{16}} \enspace .
$$
Fig.~\ref{f:thresholdplot} plots the final error probability as a function of the error $p$.  As for the Steane code, there is an unstable fixed point at $x=z=\tfrac{1}{2}$.  Unlike for previous distillation procedures $x=z=\tfrac{1}{\sqrt{2}}$ is also only an unstable fixed point, and there is an additional stable fixed point at $x=z \approx 0.62292$.
Fig.~\ref{f:acceptanceprobs} shows the acceptance probability $\tr (\ketbra{0_L}{0_L} + \ketbra{1_L}{1_L}) \rho^{\otimes 23}$.  Decoding using the Golay code is always inferior to using the Steane code.

In 
Appendix~\ref{s:moregeneralCSS},
we show that any CSS code encoding a single qubit for which logical $X$ and $Z$ operations are simply transverse $X$ and $Z$ has a fixed point at the center of each edge of the octahedron $O$.  We determine a condition on the code for this fixed point to be stable in the $H$ direction.

\section{Conclusion}

We have shown that the 7-bit Steane code pushes the error threshold for errors up to a tight value of around $\tfrac{1}{2}(1-1/\sqrt{2}) \approx 14.64\%$.  While the calculation is straightforward, it is not entirely satisfying; we do not have a good understanding of why this distillation procedure works so well.  What properties of the Steane code and the 5-qubit code allow for good distillation in the respective $H$ and $T$ directions?

It is a very interesting open question whether also the threshold in the $T$ direction is tight to the face of the octahedron $O$.
We considered dozens of small codes to try to improve the $T$-type distillation error threshold, but none could match the 5-qubit code's $17.27\%$ error threshold.  The Golay code gave a threshold of $16.12\%$, and quite a few codes have a $14.13\%$ error threshold, including the simple 2-qubit code stabilized by $ZZ$ (starting with a state along the $T$ axis it distills a state in the $xz$ plane for which the Steane code $H$-type distillation applies).
If the threshold does not meet the face of $O$, then what is the power of the intermediate region?


\bibliographystyle{bibtex/hunsrt}
\bibliography{tun}

\appendix 
\section{More general CSS codes} \label{s:moregeneralCSS}

Here we consider a general $n$-qubit CSS code encoding a single qubit, where the logical $X$ and $Z$ operations are simply transverse $X$ and $Z$.
What is the condition for $H$-type distillation to succeed for fidelities above $F_H^*$?

In the Bloch sphere, $(x,y,z)=(\tfrac{1}{2},0,\tfrac{1}{2})$ is a fixed point exactly when
\begin{multline*}
\bra{0_L} (\rho^{\otimes n} - X_L \rho^{\otimes n} X_L + X_L \rho^{\otimes n} + \rho^{\otimes n} X_L) \ket{0_L} \\ \overset{?}{=} \bra{0_L} (\rho + X_L \rho^{\otimes n} X_L) \ket{0_L} \enspace ,
\end{multline*}
or equivalently when $\bra{0_L} (2 X_L \rho^{\otimes n} X_L - \rho^{\otimes n} X_L - X_L \rho^{\otimes n}) \ket{0_L} \overset{?}{=} 0$.  For $a,b \in S$ the set of even weight classical codewords satisfying the $Z$ parity checks, let $c = a \oplus b$.  Then for $\rho$ having Bloch sphere coordinates $(x,0,x)$,
$$
\bra{a} \rho^{\otimes n} \ket{b} = \frac{1}{2^n}(1+x)^{n-\tfrac{1}{2}(\abs{a}+\abs{b}+\abs{c})}(1-x)^{\tfrac{1}{2}(\abs{a}+\abs{b}-\abs{c})}x^{\abs{c}} \enspace .
$$
Substituting $x = \tfrac{1}{2}$ to get $\tfrac{1}{2^{2n}} 3^{n-\tfrac{1}{2}(\abs{a}+\abs{b}+\abs{c})}$, making the condition for $(x,y,z)=(\tfrac{1}{2},0,\tfrac{1}{2})$ to be a fixed point just
$$
\sum_{a,b \in S} 2 \cdot 3^{\tfrac{1}{2}(\abs{a}+\abs{b}-\abs{c})} - 3^{\tfrac{1}{2}(\abs{a}-\abs{b}+\abs{c})} - 3^{\tfrac{1}{2}(-\abs{a}+\abs{b}+\abs{c})} \overset{?}{=} 0 \enspace .
$$
This is always satisfied by symmetry.  So $(\tfrac{1}{2},0,\tfrac{1}{2})$ is always a fixed point.

The point $(\tfrac{1}{2},0,\tfrac{1}{2})$ is an unstable fixed point if
$$
\frac{d}{dx} \frac{u}{2v} \vert_{x=\tfrac{1}{2}} \overset{?}{>} 1 \enspace ,
$$
where $u = \bra{0_L} (\rho^{\otimes n} - X_L \rho^{\otimes n} X_L + X_L \rho^{\otimes n} + \rho^{\otimes n} X_L) \ket{0_L}$ and $v = \bra{0_L} (\rho^{\otimes n} + X_L \rho^{\otimes n} X_L) \ket{0_L}$.  Equivalently, we ask if
$$
u'-v'-2v \overset{?}{>} 0 \enspace .
$$
With a little algebra using $\tfrac{d}{dx}\bra{a}\rho^{\otimes n}\ket{b} = \tfrac{2}{3}(n-2(\abs{a}+\abs{b}-2\abs{c})) \bra{a} \rho^{\otimes n} \ket{b}$ at $x=\tfrac{1}{2}$, this condition becomes the rather opaque inequality
$$
\sum_{a,b \in S} \left[
\begin{split}
\left(4n-1-2(\abs{a}+\abs{b}+2\abs{c})\right) 3^{\tfrac{1}{2}(\abs{a}+\abs{b}-\abs{c})} \\ - 3^{n-\tfrac{1}{2}(\abs{a}+\abs{b}+\abs{c})}
\end{split}
\right] \overset{?}{>} 0 \enspace .
$$

\section{Proof of Corollary~\ref{t:multiqubitpurestatecor}} \label{s:multiqubitpurestatecor}

Let $\ket{\psi}$ be the $n$-qubit pure state we can prepare which is not a stabilizer state, $n > 1$.  We will show that there exists a sequence of $n-1$ commuting stabilizer measurements with postselected outcomes of positive probability so that the resulting state is not a stabilizer state.  (Clifford group operations can then move the resulting state into a single qubit, for which Corollary~\ref{t:purestatecor} applies.)
The proof is by induction on $n$ the number of qubits of $\ket{\psi}$.  

Assume otherwise.  Notice that no postselected stabilizer measurement on $\ket{\psi}$ can succeed with probability 1, or else we could move that stabilizer into the last qubit, leaving an $n-1$ qubit pure state which is not a stabilizer state.  Thus
$$
\ket{\psi} = \alpha \ket{0} \ket{\psi_0} + \beta \ket{1} \ket{\psi_1} \enspace ,
$$
with $\alpha, \beta \neq 0$.
Both $\ket{\psi_0}$ and $\ket{\psi_1}$ must be stabilizer states, or we could apply the inductive assumption.
Use Clifford group operations on bits $2,\ldots,n$ to change the stabilizers of $\ket{\psi_0}$ into $\{ Z_i \}_{i=2}^n$ with all $+1$ eigenvalues; i.e. assume w.l.o.g. $\ket{\psi_0} = \ket{0^{n-1}}$.
One of the Pauli products stabilizing $\ket{\psi_1}$ must have an $X$ or $Y$ in some position -- without loss on qubit 2.  
Swap $X \leftrightarrow Y$ on qubit 2 if necessary, then use controlled-Pauli operations from qubit two to move this stabilizer into just $X_2$ (this will not affect $\ket{\psi_0}$ because its $0$ in position $2$ will not trigger the control).  By applying $Z_2$ if necessary, we may assume $\ket{\psi_1}$ is a $+1$ eigenstate of $X_2$.  Repeating the same argument gives that $\ket{\psi_1}$ has stabilizers $\{ X_i \}_{i=2}^n$ with all $+1$ eigenvalues.  Thus
$$
\ket{\psi} = \alpha \ket{0 0 0 \ldots 0} + \beta \ket{0 + + \ldots +} \enspace ,
$$
where $\ket{+} = \tfrac{1}{\sqrt{2}} (\ket{0} + \ket{1})$ the $+1$ eigenstate of $X$.  Measuring qubits $2,3,\ldots,n$ in the $Z$ eigenbasis, postselecting on $+1$ outcome, leaves the unnormalized state
$$
\alpha \ket{0} + \frac{\beta}{2^{\tfrac{n-1}{2}}} \ket{1} \enspace ,
$$
while measuring qubits $2,3,\ldots,n$ in the $X$ eigenbasis, postselecting on $+1$ outcome, leaves
$$
\frac{\alpha}{2^{\tfrac{n-1}{2}}} \ket{0} + \beta \ket{1} \enspace .
$$
Since $\alpha,\beta \neq 0$, these can't both be stabilizer states.  Thus in contradiction to our assumption, we have after all managed to reduce the state down to a single qubit pure state which is not a Pauli eigenstate.
\qed



\end{document}